\documentclass[twocolumn,english,aps,superscriptaddress,prb,floatfix,nofootinbib]{revtex4-2}
\usepackage[latin9]{inputenc}
\usepackage{color}
\usepackage{babel}
\usepackage{amsmath}
\usepackage{amssymb}
\usepackage{mathtools}
\usepackage{graphicx}
\usepackage{hyperref}

\newcommand{\tsp}{\hspace{0.4mm}}
\newcommand{\vx}{\vphantom{X}}

\newcommand{\typeN}{\textrm{N}}
\newcommand{\typeS}{\textrm{S}}

\begin{document}
\title{Stripe order in quasicrystals}

\author{Rafael M. P. Teixeira}
\affiliation{Instituto de F\'isica, Universidade de S\~ao Paulo,  05315-970, S\~ao Paulo, SP, Brazil}

\author{Eric C. Andrade}
\affiliation{Instituto de F\'isica, Universidade de S\~ao Paulo,  05315-970, S\~ao Paulo, SP, Brazil}

\begin{abstract}
We explore the emergence of magnetic order in geometrically frustrated quasiperiodic systems, focusing on the interplay between local tile symmetry and frustration-induced constraints. In particular, we study the $J_1$-$J_2$ Ising model on the two-dimensional Ammann--Beenker quasicrystal. Through large-scale Monte Carlo simulations and general arguments, we map the phase diagram of the model. For small $J_2$, a Néel phase appears, whereas a stripe phase is stable for dominant antiferromagnetic $J_2$, despite the system's lack of periodicity.  Although long-range stripe order emerges below a critical temperature, unlike in random systems, it is softened by the nucleation of competing stripe domains pinned at specific quasiperiodic sites. This behavior reveals a unique mechanism of symmetry breaking in quasiperiodic lattices, where geometric frustration and local environment effects compete to determine the magnetic ground state.  Our results show how long-range order adapts to non-periodic structures, with implications for understanding nematic phases and other broken-symmetry states in quasicrystals.
\end{abstract}

\date{\today}
\maketitle


\section{Introduction}

Quasicrystals are a unique class of ordered solids that exhibit long-range atomic arrangement without periodicity. Their distinctive diffraction patterns display sharp Bragg peaks, revealing rotational symmetries that are impossible in conventional crystals \cite{shechtman84, janssen18}. New developments have introduced novel platforms for studying quasicrystals, ranging from moir\'e-engineered systems \cite{ahn18, yao18, uri23} to optical lattices \cite{viebahn19} and arrays of artificial atoms \cite{kempkes19}, significantly broadening the scope for both experimental and theoretical exploration of these materials.

The non-periodic nature of quasicrystals presents particular challenges for understanding correlation-driven phenomena,  and a recent flurry in activity has uncovered both conventional and exotic behavior in these systems, particularly regarding superconductivity \cite{sakai17,kamiya18,araujo19,sakai19,rai20,cao20,duncan20,manna24} and magnetic properties of localized moments \cite{deguchi12,  takemura15,  andrade15,  thiem15b,  hartman16,koga17,cabrera-baez19,araujo24,tamura25}. These findings highlight the rich physics emerging from the interplay between quasiperiodic order and strong electron correlations.

Frustrated magnetic systems, where competing interactions prevent simultaneous minimization of all energies,  also host a rich variety of exotic phases and emergent phenomena \cite{lacroix11,  vojta18}.  A particularly intriguing example is the stabilization of stripe magnetic order, which spontaneously breaks both the spin-rotation symmetry and the rotational symmetry of the underlying lattice \cite{fradkin10,  fernandes19}.  While such symmetry breaking is well-studied in periodic crystals, its manifestation in quasicrystals presents an interesting problem, as they exhibit only local rotational symmetry -- such as fivefold or eightfold axes -- which is incompatible with conventional periodic order.  The interplay between this local symmetry and the symmetry-breaking stripe phase raises fundamental questions about how frustration and quasiperiodicity compete and coexist, as well as how quasiperiodic systems differ from random systems.  Studying this effect in quasicrystals not only deepens our understanding of frustrated magnetism but also reveals how long-range order can adapt -- or fail to adapt -- to systems that inherently lack translational invariance.

In this work,  we study the frustrated $J_1$-$J_2$ Ising model on the two-dimensional octagonal (Ammann--Beenker) quasicrystal \cite{socolar89,anu06,deneau89}.  For dominant antiferromagnetic $J_2$, we demonstrate the emergence of a stripe phase, analogous to the square lattice case.  Through large-scale Monte Carlo (MC) simulations, we reveal that long-range magnetic order is stable -- unlike in disordered systems \cite{andrade18, kunwar18, michel21} -- but is significantly softened by the nucleation of competing stripe domains pinned at specific local environments of the motif, i.e., the stripe phase order parameter does not saturate even at $T=0$. This behavior reflects the delicate interplay between frustration, quasiperiodic geometry, and degeneracy of the order parameter.

The paper is organized as follows: In Section~\ref{sec:j1j2},  we introduce the $J_1$-$J_2$ Ising model in the Ammann--Beenker tiling and analyze its classical phase diagram.  Section~\ref{sec:neel} introduces the MC algorithm and investigates the N\'eel order for small $J_2$ and $J_1>0$.  In Section~\ref{sec:stripe}, we examine the stripe phase, providing a detailed discussion of its stability.  Finally, in Section~\ref{sec:conclusion},  we synthesize our results, discussing their broad implications for nematic and other broken-symmetry phases in quasicrystals.

\section{\label{sec:j1j2}$J_1$-$J_2$ Ising model on the octagonal tiling}

\begin{figure}[t]
	\centering{}\includegraphics[width=1\columnwidth]{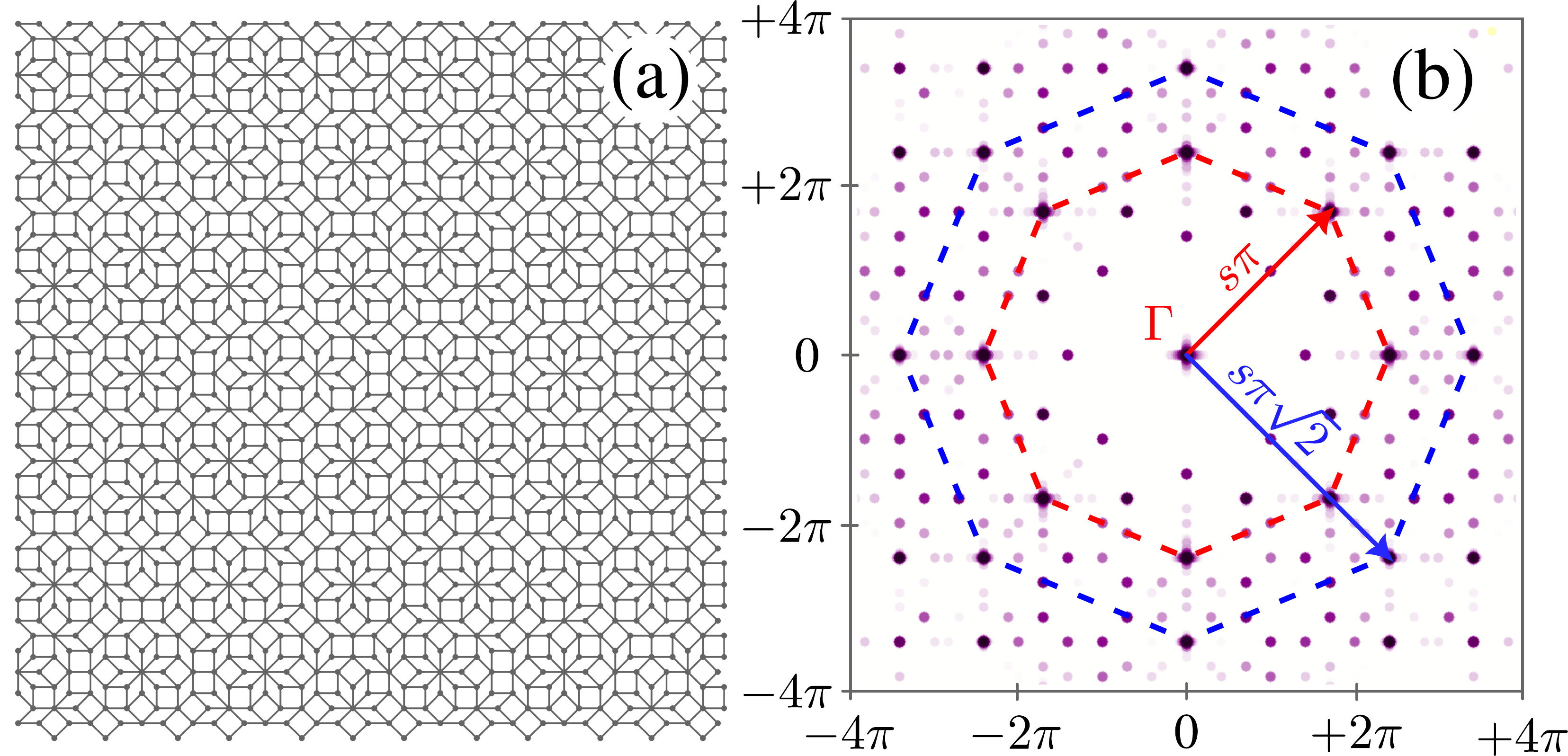}\caption{\label{fig:tiling_def}Structural properties of the
		two-dimensional Ammann--Beenker tiling.  (a) Square approximant for $N=1393$ displaying all local environments with $z=3,\ldots,8$.  (b) Fourier transform of the $N=8119$ approximant,  highlighting the eightfold rotational symmetry of the brightest Bragg peaks.  $s=1+\sqrt{2}$ is the silver ratio}
\end{figure}

We examine a simplified quasicrystal model based on the Ammann--Beenker tiling pattern. This aperiodic tiling consists of two fundamental building blocks (tiles): squares and 45$^{\circ}$ rhombuses. The structure exhibits six different atomic coordination environments, with the local coordination number $z$ ranging from 3 to 8.  Our analysis focuses on finite square approximants of increasing size,  following Ref. \cite{deneau89}.  The system size grows as $L_k=s^k$,  where $k$ is an integer defining the order of the approximant and $s=1 + \sqrt{2}$ is the silver ratio.  For simplicity,  we set the lattice spacing $a=1$.  We consider $k=3,4,5,6$,  corresponding to systems with total number of sites $N=$ $239,  1393,  8119,  47321$,  respectively (see Fig.~\ref{fig:tiling_def}(a)).  The forbidden crystallographic rotational symmetry of the approximants is evident in the static structure factor,  where an eightfold rotational symmetry encompassing the brightest Bragg peaks appears as shown in Fig.~\ref{fig:tiling_def}(b).

\begin{figure}[t]
	\centering{}\includegraphics[width=1\columnwidth]{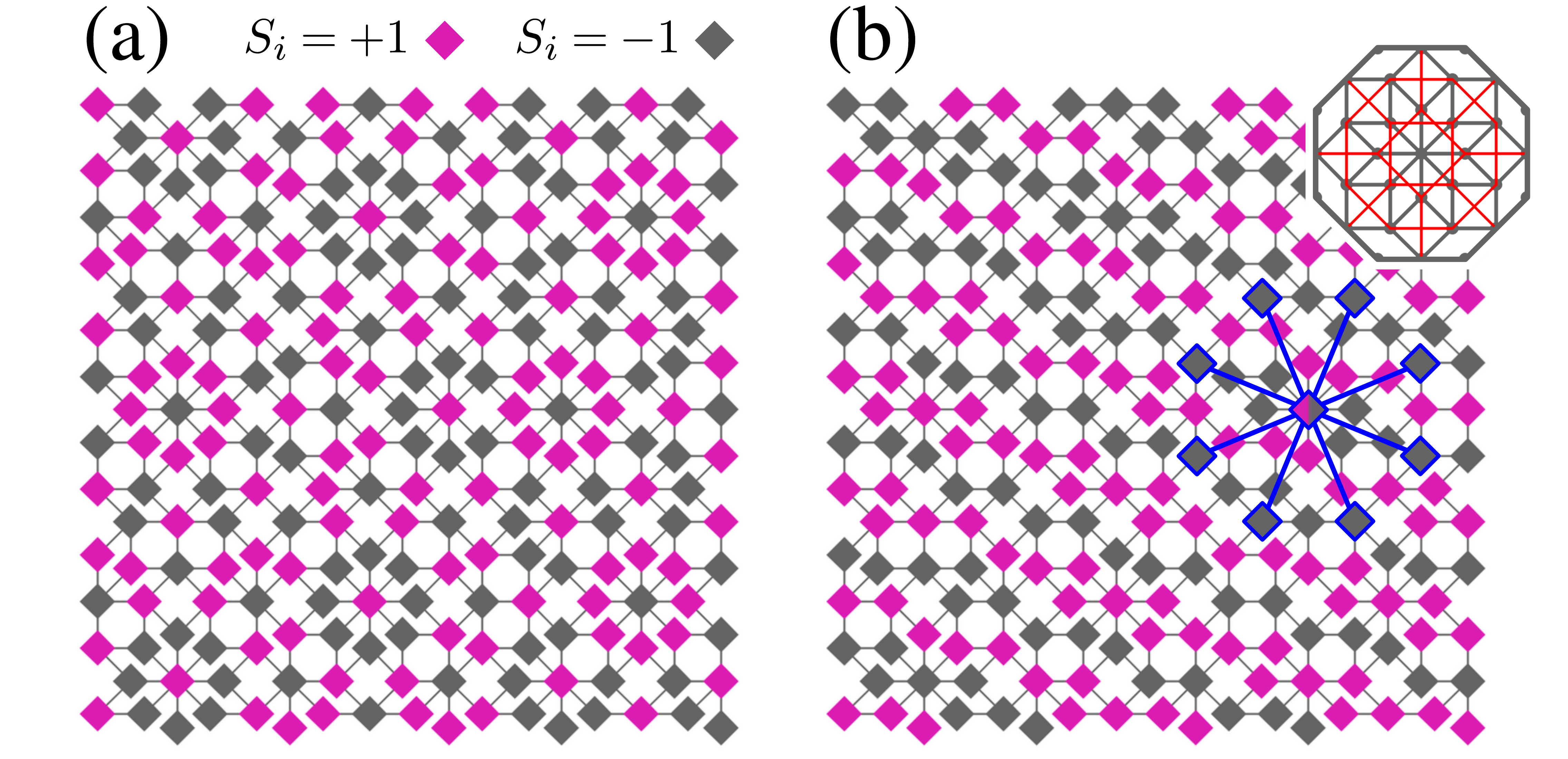}\caption{\label{fig:phases_def} Representative ground states of the antiferromagnetic $J_1$-$J_2$ Ising model on the octagonal quasicrystal.  (a) N\'eel antiferromagnetic state (the two-color coding represents the spin values $S_i=\pm 1$ at each site).  (b) One of the four possible stripe states.  Red lines highlight the coupling $J_2$ (see inset), and blue lines highlight a longer-range coupling $J_5$ at a relative distance of $d_5=2\sqrt{1/(3-s)}$, which is shown in the figure for a site with $z=8$}
\end{figure}

We are interested in the magnetism of local moments in this lattice to study the competition between quasicrystalline order and magnetic frustration.  We then study the $J_1$-$J_2$ Ising model on the approximants of the octagonal tiling
\vspace{-1mm}
\begin{equation}
	H=J_1 \sum_{\left< i,j \right>} S_i S_j + 
	J_2\!\!\sum_{\left<\left< i,j \right>\right>}\!\! S_i S_j\;,\label{eq:j1j2}
\end{equation}
where $S_i=\pm 1$ is the spin at the site $i$ and $J_{1\left(2\right)}$ is the coupling between (next-)nearest-neighbors, Fig.~\ref{fig:phases_def}(b).  We consider antiferromagnetic interactions: $J_{1} > 0$ and $J_{2} \geq 0$. While periodic boundary conditions often accelerate convergence to thermodynamic limits in crystalline systems, this advantage may not hold for quasicrystals due to their inherent variability in coordination number. Moreover, effects due to boundary conditions are rapidly suppressed with growing $L_k$ and do not affect our conclusions. We thus consider open boundary conditions and demonstrate in Sections~\ref{sec:neel} and \ref{sec:stripe} that it can successfully reproduce standard finite-size scaling behavior for thermodynamic properties,  even for nematic phases.    

The Ammann--Beenker tiling exhibits a bipartite structure, which plays a crucial role in enabling the N\'eel ordered state characterized by the formation of two interpenetrating sublattices with opposite spin orientations,  Fig.~\ref{fig:phases_def}(a). The N\'eel state is the ground state for $J_2=0$ below $T_{\typeN}/J_{1}=2.386\left(2\right)$ \cite{ledue95, repetowicz99, wessel03}.  This finding demonstrates that quasicrystalline systems can support conventional long-range magnetic order, despite their aperiodic structure, indicating that the emerging correlation length overcomes the non-trivial geometric structure of the lattice~\cite{gardner77}.

On the opposite limit,  $J_1=0$,  we expect a stripe-like state,  where each of these sublattices exhibits a  N\'eel superstructure,  analogous to what happens in the square lattice,  Fig.~\ref{fig:phases_def}(b).  Because of the local eightfold rotational symmetry,  we expect four equivalent stripe ground states,  as we discuss in detail in Section~\ref{sec:stripe}.    

\begin{figure}[t]
	\centering{}\includegraphics[width=1\columnwidth]{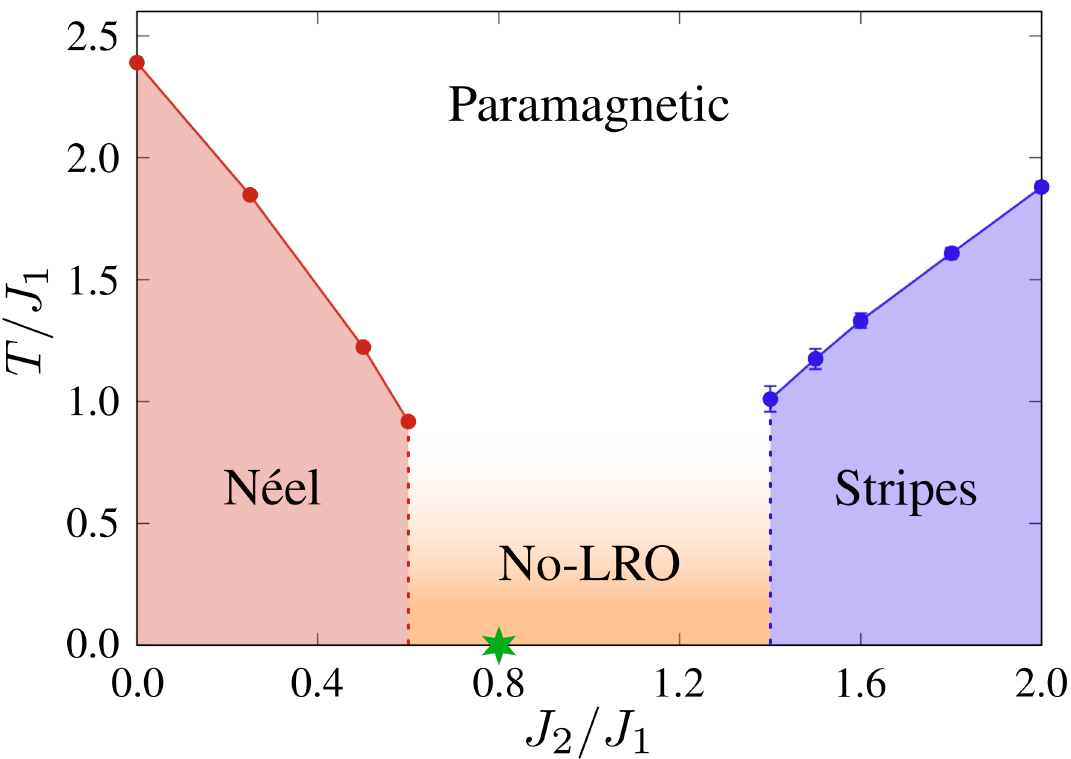}\caption{\label{fig:pd} Phase diagram of the $J_1$-$J_2$ model on the octagonal quasicrystal as a function of the temperature $T$ and next-nearest-neighbor coupling $J_2$,  with $J_1 > 0$ and $J_2 \geq 0$.  For small $J_2/J_1$,  the system displays a N\'eel ground state,  whereas for large $J_2/J_1$ we have a stripe ground state.  The transition between the two,  at $T=0$,  occurs for $J_2/J_1\simeq 0.80$,  marked by a star.  The full lines represent continuous finite-temperature transitions from the high-temperature paramagnetic to the low-temperature ordered phases.  Long-range order is observed for both small and large $J_2/J_1$ values, with the vertical dashed lines indicating their boundaries}
\end{figure}

The full phase diagram of the model as a function of $J_2/J_1$ and temperature $T$ is presented in Fig.~\ref{fig:pd}.  At $T=0$,  the transition between the N\'eel and stripe phase occurs for $J_2/J_1\simeq 0.80$,  where the energies of the ideal configuration coincide.  These ordered phases are separated from a high-temperature phase by a thermal phase transition.  Based on our simulations, we find these transitions to be continuous; however, we cannot rule out weak first-order transitions, as observed in the square lattice close to the phase boundary \cite{kalz11,jin12,kalz12}.  Our simulations reveal no long-range order in this region, even at the lowest temperatures to which the MC method can be equilibrated, due to intense competition between the two ordered phases. The boundaries of this region correspond to the dashed lines in Fig.~\ref{fig:pd}.  Moreover, domains of distinct stripe states also compete with one another within the nematic phase, thereby extending the disordered region deeper into this phase.  We will discuss these findings in detail in Sections~\ref{sec:neel} and \ref{sec:stripe}.

\section{\label{sec:neel} Monte-Carlo method and the N\'eel phase}

To investigate the Ising system governed by Eq.~\ref{eq:j1j2},  we employ classical MC simulations on a sequence of square approximants of the Ammann--Beenker quasicrystal with system sizes $N$ and open boundary conditions.  Increasing $N$ corresponds to progressively better representations of the quasicrystalline structure,  with the $N \to \infty$ limit properly recovering the true quasicrystalline properties.

Our simulation protocol combines standard single-site Metropolis updates with the parallel tempering method \cite{hukushima96,newman99} to overcome potential energy barriers and ensure proper thermalization across all temperature regimes.  This combined approach proves remarkably effective for frustrated spin systems where single-site MC methods often encounter metastability issues. The thermalization phase consists of $10^5$ MC steps per spin, ensuring the system reaches equilibrium, followed by an extensive measurement phase of $5\times10^6$ steps to obtain statistically robust averages.  Our smallest temperature is $T=0.05J_1$. 

\begin{figure}[t]
	\centering{}\includegraphics[width=1\columnwidth]{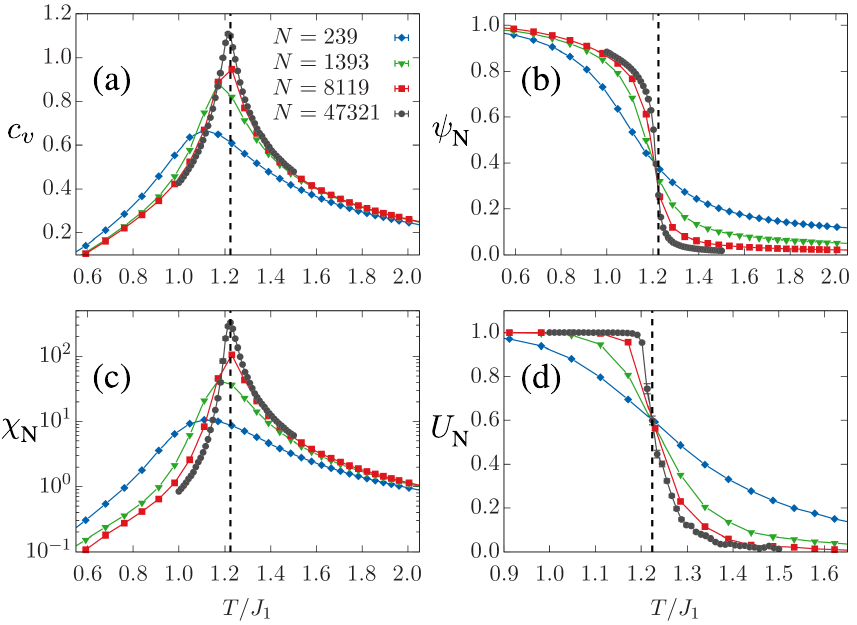}\caption{\label{fig:neel_mc} Monte Carlo results for different observables,  as a function of the temperature $T$,  characterizing the N\'eel phase for $J_2/J_1=0.5$ for different system sizes $N$.  (a) Specific heat per site $c_v$.  (b) Order parameter $\psi_{\typeN}$.  (c) Order parameter susceptibility $\chi_{\typeN}$.  (d) Binder cumulant for the order parameter $U_{\typeN}$.  The vertical dashed lines show the the transtion temperature $ T_{\typeN}/J_1 = 1.223(3)$}
\end{figure}

The bipartite structure of the tiling enables us to define a set of local phases, $\zeta_i=\pm 1$, that characterize the N\'eel ordered state,  with the different signs corresponding to distinct sublattices. In our MC simulations, we compute the N\'eel order parameter by averaging the staggered magnetization over sampled configurations,
\begin{equation}
	\psi^{\vx}_{\typeN}=\left\langle
	\frac{1}{N}\sum_{i}\zeta_{i}S_{i}\right\rangle.
\end{equation}
The susceptibility and the Binder cumulant associated with this order parameter are computed as follows, respectively:
\begin{equation}
	\chi^{\vx}_{\typeN}=\frac{N}{T}\!
	\left[\left\langle \psi^{2}_{\typeN}\right\rangle-\left\langle|\psi^{\vx}_{\typeN}|\right\rangle^{2}\right]\,,\;
	U^{\vx}_{\typeN}=\frac{3}{2}\!
	\left[1-\frac{\left\langle\psi^{4_{\vx}}_{\typeN}\right\rangle}{
		\left\langle 3\tsp\psi^{2_{\vx}}_{\typeN}\right\rangle^{2}}\right]\!.
\end{equation}
The latter is normalized, i.e., $U^{\vx}_{\typeN}\rightarrow 1(0)$ in the ordered (disordered) phase. Fig.~\ref{fig:neel_mc} displays sample MC results for $J_2/J_1=0.5$.  The specific heat peak increases slowly with the system size, indicating a continuous phase transition.  The N\'eel order parameter confirms this scenario as it saturates as $T \to 0$ for all $N$.  The order parameter susceptibility increases with the system size near $T_{\typeN}$, and the crossing point extrapolation of the Binder cumulant between different pairs of system sizes enables us to estimate $T_{\typeN}/J_1 = 1.223(4)$.  The error is estimated from the spread of these crossing points.  In selected cases, we confirmed the value of the critical temperature by a direct finite size scaling analysis of the data.

We confirm that the finite-size scaling is consistent with the Ising universality class by studying the collapse of the Binder cumulant and of the order parameter (not shown). We take $L_k$ as the linear system size for the approximant of order $k$. The results are similar to the $J_2=0$ case \cite{ledue95, repetowicz99} and other frustrated models \cite{thiem15b,  araujo24}.  We then conclude that the N\'eel phase is stable up to $J_2/J_1\simeq 0.8$,  when it gives way to the stripe phase.  The finite $T$ region around this point is difficult to access within our simulation scheme due to two significant effects: (i) fluctuations drive $T_{\typeN}$ down, and (ii) the formation of large domains of one phase inside the other.  On general grounds,  because of (ii),  one expects first-order transitions to become unstable \cite{andrade18, kunwar18, michel21, imry_ma,  aizenman_wehr,  hui89}.   However,  in a quasiperiodic system,  first-order transitions can be perturbatively stable in two dimensions \cite{barghathi14}.  Therefore, simulations at lower temperatures and larger system sizes are required to determine the fate of the ordered phases near $J_2/J_1=0.8$.

\section{\label{sec:stripe} Stripe phases in the Ammann--Beenker approximants}

\begin{figure}[t]
	\centering{}\includegraphics[width=1\columnwidth]{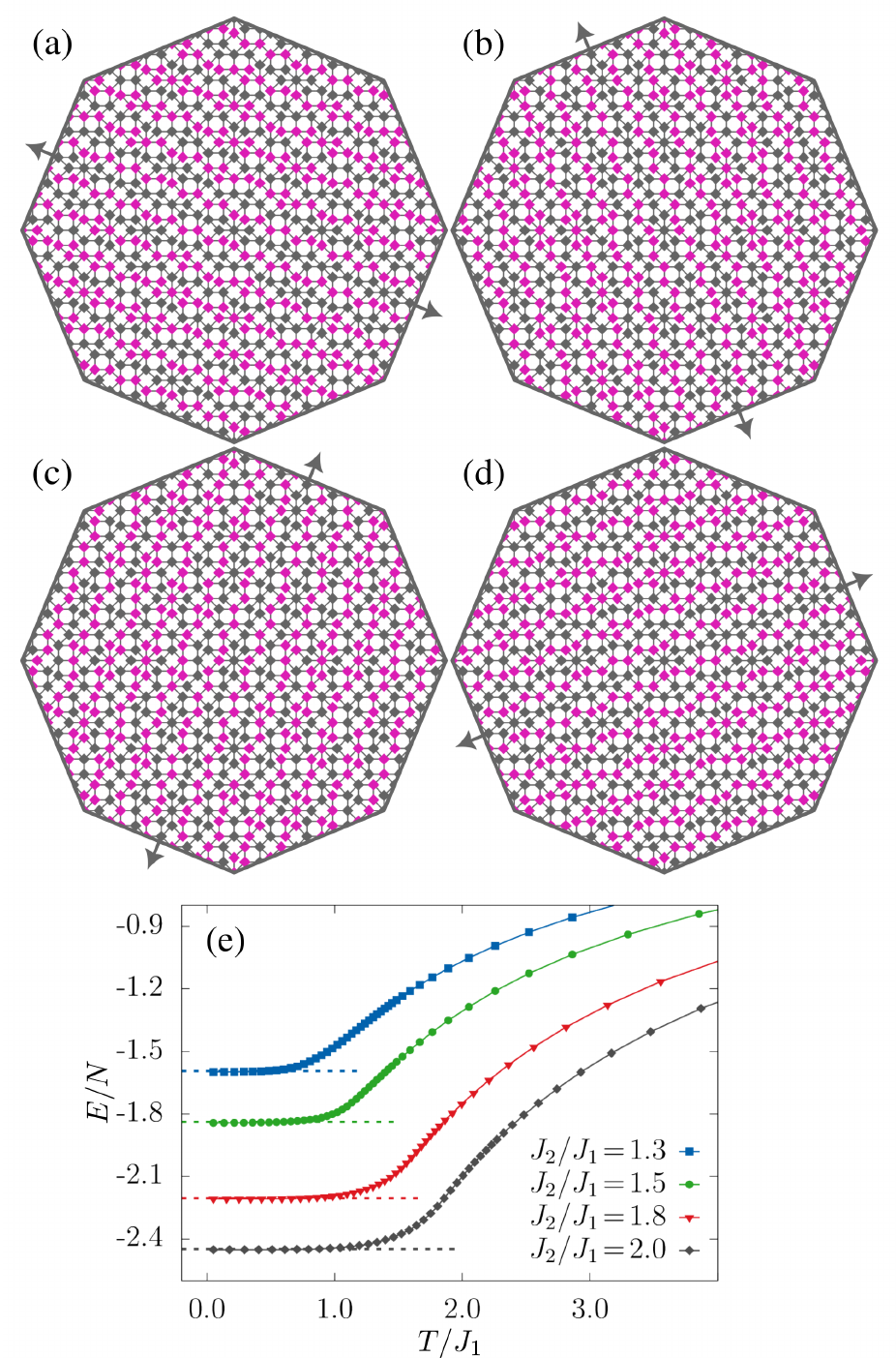}
	\caption{\label{fig:stripe_def} (a)-(d) Four different stripe patterns compatible with the Ammann--Beenker tiling $\mathcal{S}_1\,$-$\,\mathcal{S}_4$.  The color coding for the Ising spins is the same from Fig.~\ref{fig:phases_def}.  (e) Comparison between the energy,  per site,  of the Monte Carlo simulation and the energy of the perfect stripe configurations (dashed lines) for $N=8119$}  
\end{figure}

We now move to the stripe phase.  Due to the underlying symmetry of the tiling, we expect the four different stripes displayed in Fig.~\ref{fig:stripe_def}(a)-(d), which are labeled $\mathcal{S}_1\,$-$\,\mathcal{S}_4$.  In Fig.~\ref{fig:stripe_def}(e),  we compare the expected energy per spin of these ideal configurations and those of the MC simulation.  We observe that they are very close, with a relative difference of at most $0.25\%$ for $N=8119$.  This difference is due to the boundary conditions of the square approximant.  Both open and periodic conditions produce deviations with respect to the ideal stripe configurations $\mathcal{S}_1\,$-$\,\mathcal{S}_4$, which are suppressed as the system size increases, as expected.  This gives us confidence that these stripe configurations are the ground state for large $J_2/J_1$. 

\begin{figure}[t]
	\centering{}\includegraphics[width=1\columnwidth]{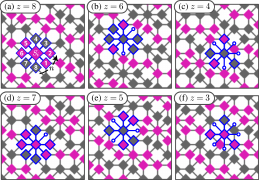}
	\caption{\label{fig:visual_guide}Six zoomed regions of the stripe state $\mathcal{S}_1$ highlighting all possible local environments on the Ammann--Beenker tiling. In (a), we show the anticlockwise labeling of nearest-neighbor spins $S_{i(n)}$ ($n=1,2,\dots,8$) used in the definition of the local order parameter in Eq.~\eqref{eq:local_OP}. For instance, in (c), we focus on a $z=4$ site with spin $S_{i}=-1$, its nearest neighbors have spin $S_{i(1,6)}=-1$ and $S_{i(4,7)}=1$, the 4 absent bonds are associated with fictitious neighbors $S_{i(2,3,5,8)}=0$ (white filled circles)}
\end{figure}

We then define the local order parameter associated with these stripe configurations.  This can be done by first considering the sites with $z=8$, which possess the complete eightfold rotational symmetry of the tiling. In Fig.~\ref{fig:visual_guide}(a), we show a schematic diagram depicting the anticlockwise labeling of the nearest-neighbor spins $S_{i(n)}$ ($n=1,2,\dots,8$) around a spin $S_i$ at some site $i$ with $z=8$. The sum of two such spins is concisely written as $\eta_{i}^{k,l}=S_{i(k)}+S_{i(l)}$. We define the complex local order parameter at this site as $M_{i}=M^{+}_i+\textrm{i}\tsp M^{-}_i$ where
\begin{equation}
	\begin{aligned}
		& M^{+}_i=\dfrac{1}{z}\,S_i\cdot\!\left(
		\eta_{i}^{1,2}-\eta_{i}^{3,4}+
		\eta_{i}^{5,6}-\eta_{i}^{7,8}\vphantom{\frac{x}{x}}\right)\\
		& M^{-}_i=\dfrac{1}{z}\,S_i\cdot\!\left(
		\eta_{i}^{2,3}-\eta_{i}^{4,5}+
		\eta_{i}^{6,7}-\eta_{i}^{8,1}\vphantom{\frac{x}{x}}\right)
	\end{aligned}\;.\label{eq:local_OP}
\end{equation}
We can extend the definition of $M_{i}$ for sites $i$ with $z<8$ by simply setting $S_{i,n}=0$ for the $8-z$ absent bonds. This is depicted in Fig. \ref{fig:visual_guide}, where one can calculate $M_{i}$ to find that $M^{+}_i=1$ in all cases, with $M^{-}_i=0$ for sites with even coordination number (a)-(c) and $|M^{-}_i|=1/z$ for sites with odd coordination number (d)-(f). The latter is finite due to an imbalance in the number of neighbors. We consider only the component that saturates (i.e., whose absolute value reaches unity) to identify the stripe state. For an ideal stripe configuration $\mathcal{S}_m$ with $m=1,2,3$ or $4$, $M_{i}$ takes the following values at all sites with even $z$: $1$, $e^{\textrm{i}\pi/2}$, $e^{\textrm{i}\pi}$ or $e^{3\textrm{i}\pi/2}$, respectively. In this way, we can locally identify the emergence of stripe configurations and map the formation of domains. Moreover, this construction allows us to define a set of local phases $\xi_{i,m}=\pm 1$ characterizing each stripe ordered state. The stripe order parameter is thus defined as
\begin{equation}
	\psi^{\vx}_{\typeS}=\left\langle\! 
	\sqrt{\sum_{m\tsp=\tsp 1}^{4}\!\left[\frac{1}{N}
		\sum_{i}\xi_{i,m}S_{i}\,\right]^{2}}\,\right\rangle,
\end{equation}
with the susceptibility and the normalized Binder cumulant associated with it given by:
\begin{equation}
	\chi^{\vx}_{\typeS}=\frac{N}{T}\!
	\left[\left\langle \psi^{2}_{\typeS}\right\rangle-\left\langle\psi^{\vx}_{\typeS}\right\rangle^{2}\right]\,,\;
	U^{\vx}_{\typeS}=2 \left[\frac{3}{2}-\frac{\left\langle\psi^{4_{\vx}}_{\typeS}\right\rangle^{\hphantom{2}}}{
		\left\langle\psi^{2_{\vx}}_{\typeS}\right\rangle^{2}}\right]\!.
\end{equation}

\begin{figure}[t]
	\centering{}\includegraphics[width=1\columnwidth]{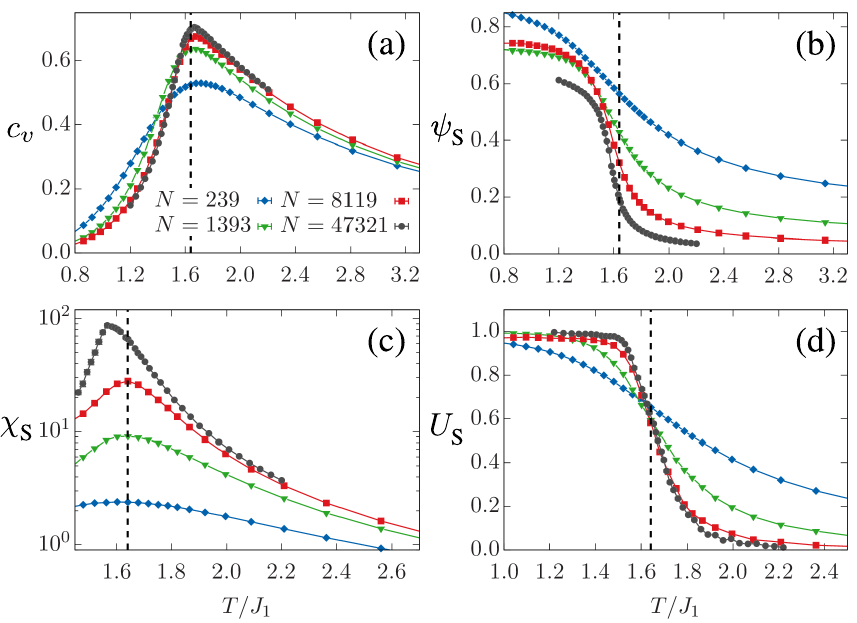}\caption{\label{fig:stripe_mc} Monte Carlo results for different observables,  as a function of the temperature $T$,  characterizing the stripe phase for $J_2/J_1=1.8$ for different system sizes $N$. (a) Specific heat per site $c_v$.  (b) Order parameter $\psi_{\typeS}$.  (c) Order parameter susceptibility $\chi_{\typeS}$.  (d) Binder cumulant for the order parameter $U_{\typeS}$.  The vertical dashed lines show the transition temperature $ T_{\typeS}/J_1 = 	1.64(5)$}  
\end{figure}

Sample results of our MC simulations considering this order parameter are shown in Fig.~\ref{fig:stripe_mc}.  The specific heat exhibits a peak that increases with system size, signaling a phase transition.  The order parameter also becomes finite at low $ T$, albeit it does not saturate due to the competition between the different stripe domains.  This competition does not disrupt long-range order, although it makes the MC sampling more challenging for the larger approximants.  These ordering tendencies can be inferred from the pronounced peak in the order parameter susceptibility and the crossing of the Binder cumulant for different system sizes.  

We were unable to extract the critical exponents for this transition.  Due to the difficulty of converging the MC runs at low temperatures,  our attempts were inconclusive.  As there are four stripe patterns,  the ordered phase breaks a $Z_4 \times Z_2$ symmetry below $T_{\typeS}$,  making it likely more complex to analyze than the already interesting case of the square lattice \cite{kalz11,jin12,kalz12}.  

The difficulties in stabilizing the stripe phase are rooted in sites with $z=8$.  For all stripe configurations, the local exchange field, $h_i=\sum_j J _ {ij} S_j$, vanishes at these sites.  This implies an ordered state that is partially disordered \cite{araujo24,  kagice24} in the sense that the spin in these sites can point in either direction even at $T=0$.  Because each of these two configurations is linked to a different stripe pattern,  the $z=8$ sites act as pinning centers capable of nucleating stripe domains other than those of the bulk.  In a random bidimensional system, this random-field-like mechanism would prevent the emergence of long-range order \cite{andrade18, kunwar18, michel21}.  In a quasicrystalline system, this is not the case \cite{barghathi14}, and the long-range order survives, albeit weakened. 

\begin{figure}[t]
	\centering{}\includegraphics[width=1\columnwidth]{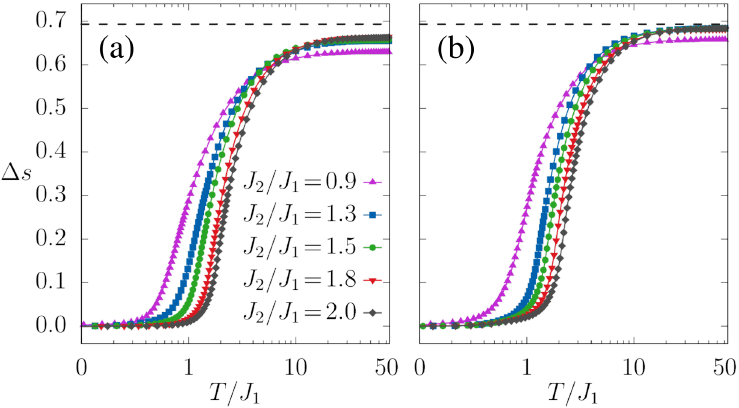}
	\caption{\label{fig:entropy} Entropy difference $\Delta s$ as a function of $T$ for $N=8119$ and different values of $J_2/J_1$ inside the stripe phase.  (a) Results for the fifth neighbor coupling $J_5=0$. (b) Results for $J_5/J_1=-0.1$}  
\end{figure}

To substantiate this scenario,  we first compute the entropy difference,  per site,  as a function of the temperature,  $\Delta s \left(T\right)=\int_0^T dT^{\prime} c_v\!\left(T^{\prime}\right)\!/T^{\prime}$,  Fig.~\ref{fig:entropy}(a).  We observe that the entropy does not saturate to $\log{2}$ at high temperature,  indicating the presence of an extensive ground state degeneracy and an associated residual entropy $s_0$.  The latter diminishes with growing $J_2$ as the stability of the stripe phase increases as we distance ourselves from the phase boundary.  From our data,  we obtain $s_0 \approx 0.03$ for the studied range of $J_2/J_1\geq 1.8$.  Since the sites with $z=8$ represent a fraction of $1/s^4$ of the total number of sites,  their contribution to the residual entropy inside the stripe phase is  $\log{2}/s^4 \approx 0.02$,  which gives a reasonable lower bound for $s_0$. The remaining entropy comes from effectively free spins located at the boundaries between different stripe configurations. 

As we have identified the sites with $z=8$ as those responsible for weakening the stripe phase, we might seek mechanisms to induce a local exchange field at these sites.  We find that the first coupling that produces a non-trivial effect comes from the fifth neighbors, located at a distance $d_5=2\sqrt{1/(3-s)}$ away, which we denote as $J_5$, as shown in Fig.~\ref{fig:phases_def}(b).  As we add a small ferromagnetic $J_5$ to the model in Eq.~\ref{eq:j1j2} for all sites in the lattice, we find that the ordered state indeed becomes more stable.  In Fig.~\ref{fig:entropy}(b), we see that $s_0 \to 0$ in the presence of $J_5$, confirming that the residual entropy is caused by the fluctuating spins at the sites with $z=8$ in the original model.  Closer to the boundary of the N\'eel phase, the residual entropy persists due to the competition between domains of the two ordered phases.

\begin{figure}[t]
	\centering{}\includegraphics[width=1\columnwidth]{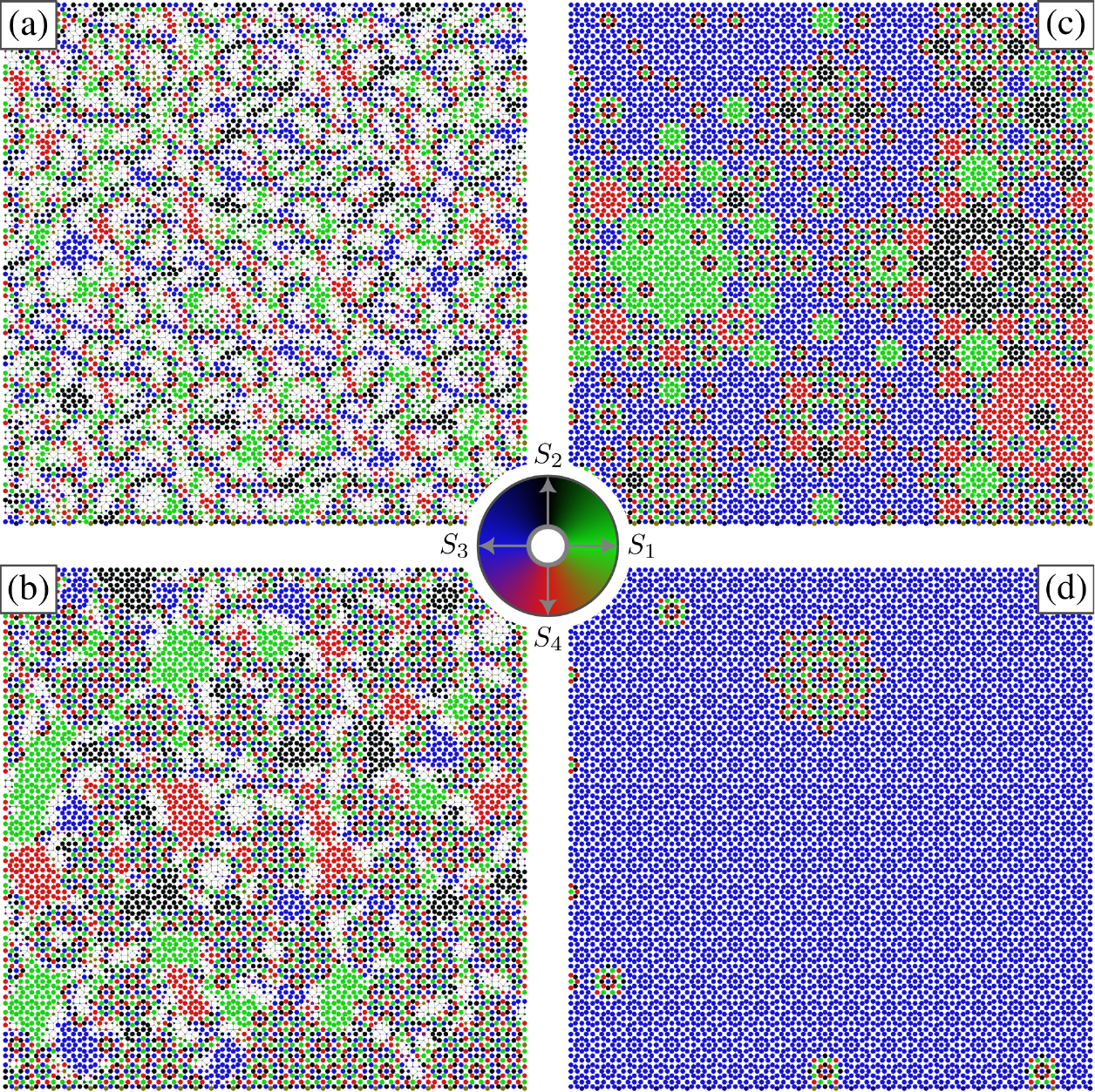}
	\caption{\label{fig:snapshots} Snapshots of the Monte Carlo configurations at low temperatures for $N=8119$.  We display the local value of the stripe order parameter across the entire lattice, with different colors indicating the four possible stripe domains, as defined by the color wheel in the center.  White means that the local stripe order parameter vanishes (or becomes very small).  (a) and (b) are snapshots for $J_2/J_1=0.9$ with $J_5/J_1=0.0$ and $J_5/J_1=-0.1$, respectively. (c) and (d) are snapshots for $J_2/J_1=2.0$ with $J_5/J_1=0.0$ and $J_5/J_1=-0.1$, respectively}
\end{figure}

We can illustrate this finding by examining the MC configuration snapshots.  We find it more convenient to consider the local value of the stripe order parameter at each site $i$ as mapped by $M_{i}$ defined in Eq.~\eqref{eq:local_OP}, and we assign a different color to stripes $\mathcal{S}_1\,$-$\,\mathcal{S}_4$ based on the phase value $\theta_i=\arg M_i$, as indicated in Fig.~\ref{fig:snapshots}.  We see that the effect of a small $J_5$ is dramatic.  For $J_2/J_1=2$, we observe that the competition between stripe domains is strongly suppressed, and a clear winning domain emerges, as shown in Figs.~\ref{fig:snapshots}(c) and (d).  Closer to the phase boundary,  $J_2/J_1=0.9$,   the stripe domains increase in size,  although no long-range order emerges,  see Figs.~\ref{fig:snapshots}(a) and (b).  This is due to competition with the N\'eel phase, which appears in the white regions between the stripe domains.  Overall, we observe that $J_5$ enhances the stability of the stripe phase by inducing a local exchange field at the site with $z=8$.  

\section{\label{sec:conclusion}Conclusion}
We have studied the antiferromagnetic $J_1$-$J_2$ Ising model on a two-dimensional quasicrystal.  Combining large-scale MC simulations and general arguments,  we find two ordered phases in the model.  For the dominant $J_1$, the system exhibits a N\'eel order, and a stripe ground state is favored for large $J_2$.  The existence of the N\'eel phase is expected as the Ammann--Beenker tiling we study is bipartite \cite{ledue95, repetowicz99, wessel03}. The stabilization of the stripe phase is more complex because it breaks a real-space symmetry of the system as it selects a preferred direction in the tiling.  While this is easily rationalized in a periodic system, as the system has global rotational symmetry, in a quasicrystalline system, one has to investigate local patterns linked to the forbidden rotational symmetry.  

In the present example,  the system displays a local eightfold rotational symmetry,  and four distinct stripe domains are stabilized.  In periodic systems,  the transition to the ordered state would select one of these states \cite{kalz11,jin12,kalz12}.   In random systems, on the other hand,  a given disorder configuration favors one of the ordered states.  Thus,  disorder acts effectively as a random field \cite{imry_ma,  aizenman_wehr,  hui89} and destroys long-range order in two dimensions \cite{andrade18, kunwar18, michel21}.  Because the critical behavior in quasiperiodic systems differs from that of random systems \cite{luck93}, the stripe phase is perturbatively stable in the problem considered \cite{barghathi14}, i.e., the long-range order survives the explicitly breaking of periodicity. We confirm the presence of the stripe phase for large $J_2$,  despite the fierce competition among the different stripe configurations leading to the formation of domains.  We also demonstrated that the longer-range coupling $J_5$ helps stabilize the stripe phase, as it suppresses the fluctuations of a particular subset of sites and favors the formation of a single stripe domain.  

It would be interesting to extend our study to the Heisenberg model, as in this case the nematic order decouples from the spin order \cite{chandra90, weber03, weber12, grison20} and one could explore distinct universality classes of the transition in a quasiperiodic system, or more exotic states like spin spirals \cite{consoli24}.  More broadly, our results indicate that nematic phases can be stable in quasiperiodic systems and that long-range orders, showing non-trivial spatial structure, can emerge on top of a quasiperiodic background. 


We thank P.  C\^onsoli and A. P.  Vieira for discussions. We acknowledge support by FAPESP (Brazil),  Grants No. 2021/06629-4,  2022/15453-0,  and 2023/06682-8.  ECA was also supported by CNPq (Brazil), Grant No. 302823/2022-0.  


%

\end{document}